\documentstyle[twoside,fleqn,espcrc2,epsf]{article}
\newcommand{\bibi}{\bibitem}
\def\a{\alpha}

\def\e{\epsilon}   
\def\g{\gamma}

\def\j{\psi}

\def\m{\mu}

\def\n{\nu}
\def\s{\sigma}    

\def\O{\Omega}

\def\jb{\overline{\j}}

\def\cD{{\cal D}}

\newcommand{\jp}{\psi^{\prime}}
\newcommand{\jbp}{\overline{\j}^{\prime}}

\newcommand{\half}{\mbox{{\normalsize $\frac{1}{2}$}} }
\newcommand{\ph}{\mbox{{\normalsize $\frac{p}{2}$}} }

\newcommand{\Tr}{\mbox{Tr\,}}
\newcommand{\ra}{\rightarrow}

\newcommand{\gm}{\gamma}

\newcommand{\vep}{\epsilon}

\newcommand{\ps}{\psi}
\newcommand{\om}{\omega}

\newcommand{\nnn}{\nonumber \\}

\newcommand{\Ps}{\Psi}

\newcommand{\psb}{\overline{\ps}}

\newcommand{\dmu}{\partial_{\mu}}

\newcommand{\hmu}{\hat{\mu}}

\newcommand{\be}{\begin{equation}}
\newcommand{\ee}{\end{equation}}
\newcommand{\bea}{\begin{eqnarray}}
\newcommand{\eea}{\end{eqnarray}}
\newcommand{\eq}{\ref}
\newcommand{\beq}{\begin{equation}}
\newcommand{\eeq}{\end{equation}}
\newcommand{\cc}{\cite}
\newcommand{\lb}{\label}

\def \3{\ss}

\newcommand{\AmS}{{\protect\the\textfont2
  A\kern-.1667em\lower.5ex\hbox{M}\kern-.125emS}}

\title{Can baryogenesis occur on the lattice?}

\author{Wolfgang Bock\address{Dept. of Physics, University of
                              California, San Diego, Gilman
                              Dr.~0319, La Jolla, CA 92093-0319,
                              USA}\thanks{talk presented by W. Bock}
        and James E. Hetrick\address{Physics Department,
                              University of Arizona, Tucson
                              AZ 85721, USA}
       }

\begin{document}

\begin{abstract}
We examine the question of how baryogenesis can occur in
lattice models of the Standard Model where there is a global $U(1)$
symmetry which is accompanied by an exactly conserved fermion number.
We demonstrate that fermion creation and annihilation can occur
in these models
{\em despite} this exact fermion number conservation,
by explicitly computing the
spectral flow of the hamiltonian in the two dimensional U(1) axial model
with Wilson fermions.
For comparison we also study the closely related Schwinger model where
a similar mechanism
gives rise to anomalous particle creation and annihilation.
\end{abstract}

\maketitle

\section{INTRODUCTION}
It is well known that fermion number is not conserved in the Standard
Model (StM) due to the anomaly in the baryon and lepton currents which
arises from the rich vacuum structure of non-abelian gauge
theories. Topologically distinct vacua which are characterized by
their integer Chern class $n$ are separated by a potential barrier
whose minimal height is given by the so-called sphaleron energy,
$E_{{\rm sph}}$. Transitions from one vacuum to another are
accompanied by a change in baryon and lepton numbers, $\Delta
Q_{B,L}=N_G \Delta n$, with $N_G$ the number of
generations. It was shown in
ref.~\cc{Ho76} that the tunneling transition rate at zero temperature
is enormously suppressed by the small exponential factor
$e^{-4\pi/\a_W} \approx e^{-150}$; $a_W$ being the weak fine structure
constant. As was first pointed out in ref.~\cc{KuRu85} though, the
transition rate gets substantially enhanced at large temperatures,
since the system can travel classically from one vacuum to another by
thermal excitation instead of having to penetrate the barrier by
tunneling. This happens when the thermal energy approaches the
sphaleron energy and the transition rate is then proportional to the
Boltzmann factor $\exp(-E_{{\rm sph}}(T)/T)$.

The existing semiclassical calculations of the transition rate are
based on certain assumptions and it is important to check the results
by a calculation from the first principles and the only possibility at
present to address such a non-perturbative problem is to use lattice
Monte Carlo calculations. Some first attempts in this direction
have been performed in the SU(2) gauge-Higgs sector of the StM
\cc{AmAs} and in the two dimensional abelian Higgs model \cc{GRS}.
However these simulations must be extended to the full StM including
fermions. A potential problem however is that a
satisfactory formulation of the StM is still lacking. On the lattice
each fermion is accompanied by spurious species doublers of opposite
chirality which render chiral gauge theories vector-like. In the past
three years  most models, which have been proposed so far to
overcome this phenomenon, have been shown to fail. The situation looks
even worse with regard to the issue of fermion non-conservation. It
has been argued in ref.~\cc{Ba91} that currently existing models are
unable to reproduce the fermion number non-conservation of the StM,
since the action which in general is of the form
\be
S = -\sum_{x,y}\psb_x D(A)_{xy} \psi_y
\ee
and also the measure in  the path integral are
invariant under a global U(1)
symmetry $\psi_x \ra \exp (i\a) \psi_x$,
$\psb_x \ra \psb_x \exp (-i\a)$. One therefore
expects this exact global U(1) symmetry to be accompanied by an exactly
conserved fermion number, and hence no baryogenesis on the lattice
when fermions are included in the above simulations.

We shall outline in this contribution how anomalous fermion creation
and annihilation can occur on the lattice {\em despite} the exact
fermion number conservation. For the sake of simplicity we have
illustrated our reasoning for the case of a two dimensional
axial model with Wilson fermions, however we see no reason that
the results would not generalize to four dimensions, or to models with
Higgs fields. Most of this work has been previously presented in
ref.~\cc{BoHe94} using  staggered fermions.
\section{AXIAL AND VECTOR QED$_2$}
The massless two-dimensional axial model is given by
the following (real time) continuum action
\bea
S\!\!&=&\!\! -\int d^2 x \frac{1}{4e^2} F_{\mu\nu}F^{\mu\nu} + S_F
\;, \lb{FSQ} \\
S_F\!\! &=&\!\! -\int d^2 x  \jb \g^{\mu}
(\dmu + i A_{\m}\g_5) \j \;,\lb{AXQED}
\eea
where $\g^1=\g_1=\s_1$, $\g^0 = -\g_0 = -i\s_2$, $\g_5 = \s_3$.
We take $A_{\mu}$ to be an external gauge field; only the fermion
fields are quantized. The model is invariant under local axial gauge
transformations: $\j(x) \ra \exp( i \om(x) \gm_5)
\j(x)$, $\jb(x)\ra \jb(x) \exp( i \om(x) \gm_5)$, $A_{\m}(x)
\ra A_{\m}(x) + \dmu \om(x)$, with gauge current $j^{\mu}_5 = i \jb
\g^{\mu} \g_5 \j$.
The action is furthermore invariant under the global U(1) symmetry
$\j(x) \ra \exp(i \a) \j(x)$, $\jb(x) \ra \jb(x) \exp(-i
\a)$, with a corresponding globally conserved vector current
$j^{\mu} = i \jb \g^{\mu}\j$. According to standard lore the (local)
gauge current has to be anomaly free whereas the vector current
becomes anomalous, $\dmu j^{\m}= -2 q$ where $q=\frac{1}{4\pi} \e^{\m
\n} F_{\m \n} \equiv \dmu C^{\m}$ denotes the topological charge
density and $C^{\m}(x)=(1/2\pi) \e^{\m \n} A_{\n}(x)$ is the
Chern-Simons current. The axial and vector charges are defined
as $Q_5=\int dx^1 j_5^0(x)$ and $Q=\int dx^1 j^0(x)$, and the
Chern-Simons number as $C=\int dx^1 C^0(x)$. From the current
divergence equation one sees that
\be
Q(t)-Q(0)=-2 (C(t)-C(0)) \lb{QC}
\ee
which relates a time dependent change in the Chern-Simons number to a
time dependent change in the fermion number. Thus a change in the
topological charge of the gauge field $\Delta C=-1$ gives rise to the
creation of a left and right-handed fermion. The aim of this
contribution is to investigate whether, in lattice models which have
exact global U(1) symmetry (i.e. the vector current is
divergence free in contrast to the above continuum example), a change in
$C$ due to a sphaleron transition can still give rise to a change in
the fermion number in accordance with eq.~(\eq{QC}).

A nice feature of the two-dimensional axial model is that it is
equivalent to the massless Schwinger model (QED$_2$) which is well
studied. By performing a charge conjugation transformation on only the
right (or left)-handed fermion fields in eq.~(\eq{AXQED}), i.e.  $\j_R
=(\jbp_R {\cal C})^T$, $\jb_R = - ({\cal C}^{\dagger} \jp_R)^T$, $\j_L
=\jbp_L$, $\jb_L = \jp_L$ one finds that eq.~(\eq{AXQED}) turns into
\be
S_f^\prime = -\int d^2 x \jbp \g^{\mu}(\dmu - i
A_{\mu})\jp\;, \lb{VCQED}
\ee
which is the action of vector QED with massless fermions in two
dimensions. We put a prime in the sequel to label quantities of the
vector model. Under the same charge conjugation it can be shown that
the axial current turns into the vector current, $j^\m_5 \ra -i
\jbp \g^{\mu} \jp = -j'^{\mu}$, which is now the divergence free gauge
current, $\dmu j^\m =0$. The global vector current becomes the global
axial current $j^{\mu} \ra -i \jbp \g^{\mu} \g_5 \jp =
-j_5^{\prime \mu}$ and is anomalous, $\dmu j_5^\m=2 q(x)$.
Correspondingly $Q \ra -Q_5^{\prime}$, $Q_5 \ra
-Q^{\prime}$ for the charges. From the divergence relation for
$j_5^\m$ we can derive the following formula
\be
Q^{\prime}_5(t)-Q^{\prime}_5(0)= 2 (C(t)-C(0)) \;. \lb{Q5C} \\
\ee
A change in Cern-Simons number $\Delta C=-1$ gives rise to the
creation of a left-handed fermion and a right-handed antifermion. It
is therefore instructive to study the spectral flow of the vector
model in conjunction to the axial model.

In our earlier publication we used staggered fermions to transcribe
the two models to the lattice \cc{BoHe94}. Here we shall use instead
Wilson fermions which may be somewhat more transparent. The
euclidean lattice path integral for external gauge field configuration
is then given by
\be
Z = \int (\prod_x \cD \psb_x \cD\psi_x)\, \exp [-(S_F+S_W)], \lb{PINT}
\ee
with
\be
S^{\prime}_F=\half \sum_{\m x} ( \jbp_x \g_\m U_{\m x} \jp_{x+\hmu}
-   \jbp_{x+\hmu} \g_\m U_{\m x}^* \jp_x ) \lb{LV}
\ee
for the vector QED$_2$ and
\bea
S_F\!\!\!&=&\!\!\! \half \sum_{\m x}
( \psi_x \g_\m  (U_{\m x} P_L+U_{\m x}^*) P_R)
 \psi_{x+\hmu}  \nnn
\!\!\!&&\!\!\! -\psi_{x+\hmu} \g_\m
(U_{\m x}^*P_L+U_{\m x} P_R) \psi_x )\;. \lb{LAV}
\eea
for the axial model.
We use lattice units in which the lattice distance $a=1$.
$U_{\m x}=\exp(-iA_{\m x})$ is the lattice gauge field
and $P_{R,L}=\half(1\pm \g_5)$ are the
chiral projectors.
We have added in (\eq{PINT}) a Wilson term of the form
\bea
S_W\!\!\!&=&\!\!\! r \sum_{\m x} \left(\jbp_x \jp_x
-\half \left\{ \jbp_x U_{\m x} \jp_{x+\hmu} \right. \right.\nnn
\!\!\!&&\!\!\!  \left. \left. +  \jbp_x U_{\m x}^* \jp_{x+\hmu} \right\}
\right) \;. \lb{WIL}
\eea
(and for the axial model with $\jp_x \ra \j_x$)
which for external gauge fields leaves
the physical fermion at momentum $(0,0)$ massless and removes
the three species doublers, that are located in the two
dimensional Brillouin zone at momenta $(0,\pi)$, $(\pi,0)$ and
$(\pi,\pi)$, from the spectrum by giving them a mass of the order
of the cutoff.

The vector model is invariant under local vector gauge transformations
$\psb_x \ra \psb_x \O_x^*$, $\psi_x \ra \O_x \psi_x $, $U_{\m x} \ra
\O_{x} U_{\m x} \O^*_{x+\hmu}$. The Wilson term however breaks the
global chiral invariance but it has been shown ref.~\cc{KaSm} that the
violation of this global symmetry turns, in the scaling region, into
the usual chiral anomaly. We investigate in the following how this
manifests itself in the spectral flow of the hamiltonian when the
gauge field is slowly changed from one vacuum configuration to a
topologically distinct one.
%
%
%
\begin{figure}[htb]
\vspace{0.8cm}
\centerline{
 \epsfysize=7.5cm
 \epsffile{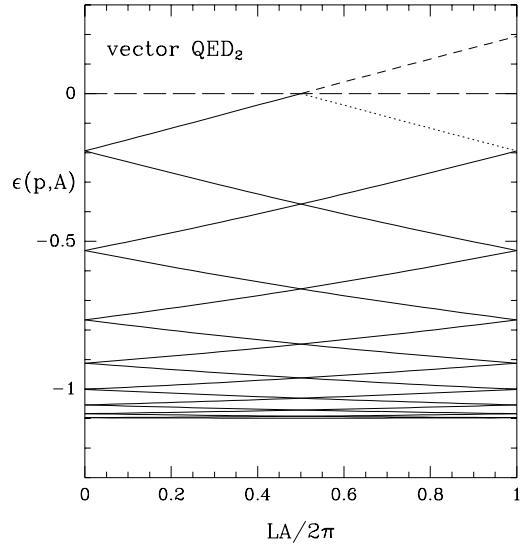}
}
\vspace{-1.2cm}
\caption{ \noindent {\em Spectral flow for vector QED$_2$.
}}

\label{FIG1}
\end{figure}
An earlier analysis of the spectral flow in vector QED$_2$ with Wilson
fermions has been presented in ref.~\cc{AmGr}.

The path integral of
the axial model is invariant under the unwanted global U(1)
symmetry. There however the ({\em local}) chiral gauge invariance
is broken by the Wilson  term. It has been shown in
ref.~\cc{BoHe94} that gauge invariance can be restored in the
perturbative regime by adding a mass counterterm for the gauge field
to the action, however a complete chirally gauge invariant lattice
regulator is still lacking and is a fundamental problem of lattice
gauge theory. As mentioned earlier most of the very promising
proposals for a lattice formulation of chiral gauge theories have been
shown to fail. Much work remains to be done in this direction,
and our present study of the subtleties of anomalous fermion
production seems to be a necessary step, which is also of interest in
its own right.
\section{INVESTIGATION OF THE SPECTRAL FLOW}
In this section we study the spectrum of the fermionic
hamiltonian in an external gauge field,
which starts out as a vacuum gauge field, goes through
a sphaleron like configuration and ends up as a different
vacuum gauge field configuration that is related to the
first one by  a topologically non-trivial (`large') gauge
transformation. We shall only consider
gauge fields which vary very slowly in
time so that we can make use of the adiabatic
theorem and determine the spectrum of the hamiltonian by
continuity. From the euclidean path integral one can derive
the transfer operator $T$, $Z=\Tr T^{N}$, where $N$ is the number
of time slices in the lattice. The hamiltonian is defined in terms of $T$
by $T=\exp (-H)$. We shall consider in the following the simple
gauge field $A_{1 x}=A(t)$, $A_{0 x}=0$ (i.e. $U_0=1$). Then the time
dependence of the hamiltonian is expressed as $A$ dependence. The
Chern-Simons number is given by $C=-LA/2\pi$.
With this choice of the gauge field configuration we  can diagonalize the
transfer operator explicitly and then determine from its eigenvalues
the spectrum of the associated hamiltonian. We shall furthermore
use anti-periodic boundary conditions for the fermions in space
which implies that the spatial momenta  are given by
$p=(n-\half)2\pi/L$, $n=-\frac{L}{2}+1,\ldots,\frac{L}{2}$ with
$L$ denoting the extent of the lattice in space direction.
\subsection{Spectral flow in vector QED$_2$}
In the vector model we derived for
the enery spectrum $\vep(p,A)$ of the hamiltonian the following expression
\bea
\vep(p,A)\!\!\!&=&\!\!\! \vep_0(p-A)  \lb{EVEC1} \\
\cosh \vep_0(p)\!\!\!&=&\!\!\! \frac{1+4 \sin^2\ph}{1+2 \sin^2\ph}
\;, \lb{EVEC2}
\eea
where $\vep_0(p)$ are energies for free Wilson fermions. For momenta $p$
and gauge fields $A$ which are small compared to the cutoff $\pi$
(\eq{EVEC1}) reduces to the continuum expression
$\vep(p,A)= \pm (p-A)$ for linearly spaced modes $p$.
The physical vacuum (Dirac sea) is the state of lowest energy
in which all levels with negative energy
are occupied. The spectrum
of the hamiltonian has been plotted in fig.~1 as a function of $LA/2\pi
=-C$ using a lattice of size $L=16$.  For clarity we have displayed
only the states in and close to the surface of the Dirac sea.
The $L$ states move upwards whereas the $R$ state move
downwards. Midway between $A=0$ and $A=2\pi/L$ the Dirac sea loses an
$L$ state and gains an $R$ state (coming from the Dirac sky).

Consider now the flow of the physical state $|\Psi, A\rangle $ which
at $A=0$ starts out as the vacuum state $|0,A\rangle$. The occupied
levels of that state are represented by the solid and dashed lines.
Now assign the quantum numbers $Q^{\prime}$ and $Q^{\prime}_5$
to this state. The
axial charge $Q^{\prime}_5$ is given by the number of $R$ states minus the
number of $L$ states and the fermion number $Q^{\prime}$
is given by the sum of
$L$ and $R$ states minus half the total number of states. From this it
is clear that the charges $Q{\prime}_{5,\Psi}$
and $Q^{\prime}_{\Psi}$ of the physical
state are both zero in the initial state at $A=0$ and the final state
at $A=2\pi/L$.  The crucial observation however is that fermions are
excitations relative to the vacuum, which implies that the change in
the axial charge is given by
\bea
\Delta Q^{\prime}_5 \!\!\!&\equiv&\!\!\!
(Q^{\prime}_{5,\Ps}-Q^{\prime}_{5,0})_{\rm final} -
(Q^{\prime}_{5,\Ps}-Q^{\prime}_{5,0})_{\rm initial}\;.
\nnn
   && \lb{Q5FI}
\eea
The axial charge of the ground state $Q^{\prime}_{5,0}$ changes
by two  units when changing $C$ from $0$ to $-1$ since it
looses an $L$ state and gains an $R$ state, i.e.
$(Q^{\prime}_{5,0})_{{\rm initial}}=0$,
but $(Q^{\prime}_{5,0})_{{\rm final}}=-2$.
The result is that $\Delta Q^{\prime}_5=-2$  which is in concordance with
eq.~(\eq{Q5C}) since $\Delta C=-1$ in our case.
Similarly it can be shown also that $\Delta Q^{\prime}=0$.
\begin{figure}[htb]
\vspace{-0.1cm}
\centerline{
 \epsfysize=7.5cm
 \epsffile{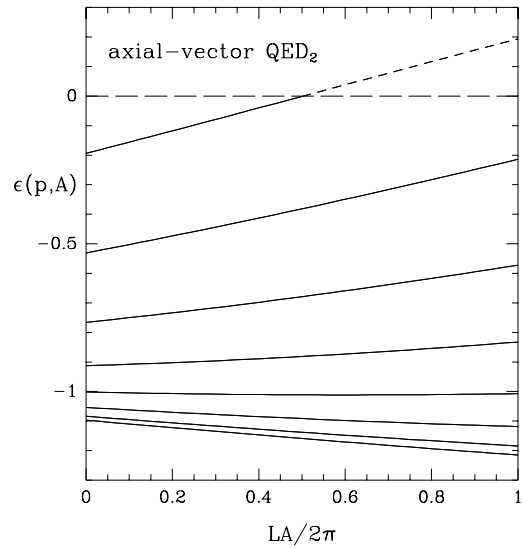}
}
\vspace{-1.2cm}
\caption{ \noindent {\em Spectral flow for axial QED$_2$.
}}
\label{FIG2}
\end{figure}
Vector QED$_2$ is gauge invariant and the vacua at $A=0$ and
$A=2\pi/L$ are related by a large gauge transformation. In Hilbert
state, these gauge transformations induce unitary transformations
on the hamiltonian which implies that its spectra coincide
at $A=0$ and $A=2\pi/L$. We also mention
that there is a peculiarity about the lowest state in the spectrum
which corresponds to the largest momentum. It
starts out as an $L$ state (moving downwards) and ends
up as an $R$ state (moving upwards). This change of chirality
appears to be possible because the Wilson term mixes $L$ and $R$ states.
The chirality flip occurs however only for the largest
momentum and hence has to be regarded as a lattice artefact.
\subsection{Spectral flow in axial QED$_2$}
The energy spectrum in the axial model is given
by the formula
\bea
\vep(p,A)\!\!\!&=&\!\!\! \ln
\sqrt{ \frac{1+\sin A \cos p +2 \sin^2 \ph}
            {1-\sin A \cos p +2 \sin^2 \ph}  } \;\;\;\;+ \nnn
&&\!\!\!\!\!\!\!\!\!\!\!\!\!\!\!\!\!\!\!\!\!\!
\!\!\!\!\!\!\!\!\!\cosh^{-1}
\frac{1+4\sin^2 \ph-\half \sin^2 A}
     { \sqrt{(1+2\sin^2 \ph)^2-\sin^2A (1-2\sin^2 \ph)} }\;. \nnn
&& \lb{EAX2}
\eea
where we have dropped the gauge fields from
the Wilson term in eq.~(\eq{WIL}). Both the
Wilson terms (with and without $U$ fields)  break
the (local) axial gauge invariance.
Eq.~(\eq{EAX2}) reduces for momenta $p$
and gauge fields $A$, which are small compared to the cutoff,
to the continuum expression
$\vep (p,A)=\pm p+A$ for linearly spaced modes $p$.
The energies $\vep(p,A)$ have been displayed in fig.~2, again
plotted as a function of $LA/2\pi$. Each state is doubly degenerate
since it involves an $R$ and an $L$ state.
Consider now the flow of
the state $|\Psi,A\rangle$ which starts out at $A=0$
as the vacuum state $|0,A\rangle$. Like in
the vector model, the state $|\Psi,A\rangle$ does
not change quantum numbers. i.e.
$(Q_{\Ps})_{\rm initial}=0$,
$(Q_{\Ps})_{\rm final}=0$.
The vacuum  however
changes fermion number since it looses two states half way
between $A=0$ and $A=2\pi/L$ where an $L$ and an $R$ state cross
the surface of the Dirac sea. For the vacuum state we find
$(Q_{0})_{\rm initial}=0$, $(Q_{0})_{\rm final}=-2$ and  hence
\bea
\Delta Q \!\!\!&\equiv&\!\!\!
(Q_{\Ps}-Q_{0})_{\rm final} - (Q_{\Ps}-Q_{0})_{\rm initial}\;.
\nnn
         \!\!\!&=&\!\!\!+2 \;. \lb{QFI}
\eea
which agrees
with the continuum result (\eq{QC}). In contrast to the vector
case the spectra at $A=0$ and $A=2\pi/L$ are
not identical which is due to
lack of gauge invariance. For modes near the maximal
negative or positive momenta we see clearly regularization effects,
while the energy of modes near the surface of the Dirac sea return
roughly to their original energies.
Fig.~2 shows that the slope of the lines decreases
when increasing $|p|$ and
finally even becomes negative. We should stress that such an
effect has not been observed with  staggered fermions showing that
the regularization effects are smaller for staggered fermions. \\

\noindent {\bf Summary:}
We have shown based on the investigation of the spectral flow in the
external field approximation, that anomalous fermion creation and
annihilation can occur, even when fermion number is conserved. The
crucial observation is that fermions are excitations relative to the
vacuum. The global U(1) symmetry prohibits a state from changing its
fermion number, however nothing prevents the ground state from doing
so.
\\

\noindent {\bf Acknowledgements:} The work on spectral flow with
staggered fermions has been done in collaboration with Jan Smit.
We thank him for many interesting discussions.
This research was supported by
the DOE under contract DE-FG03-90ER40546.

\end{document}